# Bifurcation analysis of the Keynesian cross model

Xinyu Li

University of Washington

Abstract

This study rigorously investigates the Keynesian cross model of a national economy with a focus on the dynamic relationship between government spending and economic equilibrium. The model consists of two ordinary differential equations regarding the rate of change of national income and the rate of consumer spending. Three dynamic relationships between national income and government spending are studied. This study aims to classify the stabilities of equilibrium states for the economy by discussing different cases of government spending. Furthermore, the implication of government spending on the national economy is investigated based on phase portraits and bifurcation analysis of the dynamical system in each scenario.

## 1. Introduction

The original Keynesian model was first proposed to explain the persistent unemployment after the Great depression [1]. Since then, it has evolved into many modified models such as the neoclassical Keynesian model [2], the post Keynesian model [3], and the IS-LM model [4]. This paper focuses on one modification of the original Keynesian model – the Keynesian cross model [5].

In this paper, a simple model of a national economy is studied based on the Keynesian cross model. The model in this paper aims to quantify the dynamic relationship between national income and consumer spending, with special attention to the impact of government spending on economic stabilization. This paper applies bifurcation analysis, a technique commonly used in dynamical systems in mathematics, to investigate the dynamics between national income and consumer spending. The rest of the paper is organized as follows: Section 2 describes the methodology, variables, and the mathematics model used in this study; Results of the economic equilibrium are discussed in Section 2. Finally, Section 4 concludes the paper.

## 2. Method

The Keynesian cross model builds upon two ordinary differential equations [6]:

$$\dot{I} = I - \alpha \cdot C \qquad (1)$$

$$\dot{C} = \beta \cdot (I - C - G) \qquad (2)$$

where $C \geq 0$ is the rate of consumer spending, $I \geq 0$ is the national income, and $G \geq 0$ is the rate of government spending. The parameters $\alpha$ and $\beta$ satisfy $1 < \alpha < \infty, 1 \leq \beta < \infty$. Three relations between government spending and national income are discussed in the following subsections.

### 2.1. G is constant

Consider a model consisting of equations (1) and (2) along with a constant government spending G. To determine the equilibrium state for this model, I find the point where $= \dot{C} = 0$. Rearranging terms, I obtain the following equilibrium:

$$(I, C) = \left(\frac{\alpha \cdot G}{\alpha - 1}, \frac{G}{\alpha - 1}\right) \tag{3}$$

In order to calculate the stability of this fixed point, I compute the Jacobian matrix and eigenvalues:

$$J = \begin{bmatrix} \frac{\partial \dot{I}}{\partial I} & \frac{\partial \dot{I}}{\partial C} \\ \frac{\partial \dot{C}}{\partial I} & \frac{\partial \dot{C}}{\partial C} \end{bmatrix} \tag{4}$$

$$\lambda_{1,2} = \frac{(1-\beta) \pm \sqrt{\beta^2 + 2\beta + 1 - 4\alpha\beta}}{2} \tag{5}$$

The eigenvalue expression shows that, as long as $\beta > 1$, there exist two fixed points if $\Delta = \sqrt{\beta^2 + 2\beta + 1 - 4\alpha\beta} > 0$, one fixed point if $\Delta = \sqrt{\beta^2 + 2\beta + 1 - 4\alpha\beta} = 0$, and no fixed point if $\Delta = \sqrt{\beta^2 + 2\beta + 1 - 4\alpha\beta} < 0$. In the first case $\Delta > 0$, $\lambda_1$ and $\lambda_2$ have opposite signs. Accordingly, there exists a stable spiral when $\alpha > \frac{(\beta-1)^2}{4\beta} + 1$. A stable star exists in case 2 when $\alpha = \frac{(\beta-1)^2}{4\beta} + 1$ and a stable node exists in case 3 when $\alpha < \frac{(\beta-1)^2}{4\beta} + 1$. When $\beta = 1$, the eigenvalues are imaginary and there exists a center.

Figure 1 shows three phase portraits of three different $\alpha$ values when $\beta > 1$. Figure 2 displays the phase portrait of the center when $\beta = 1$.

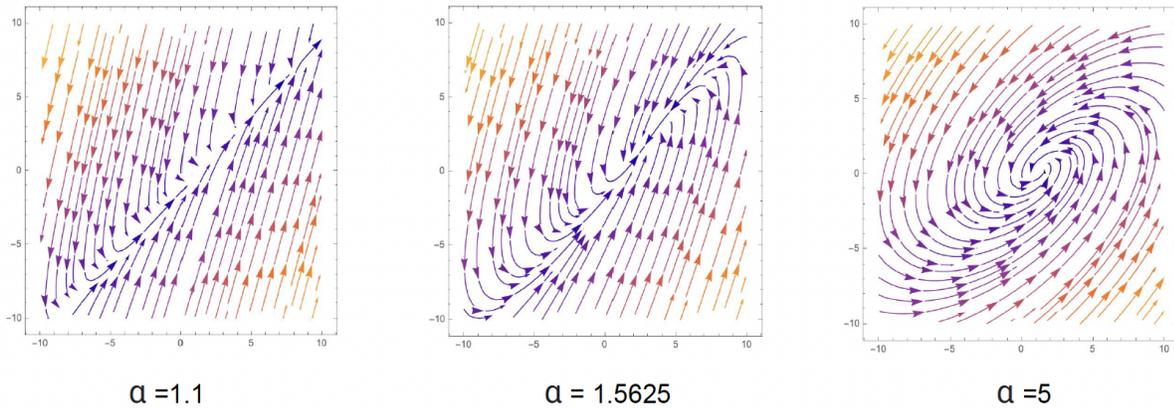

**Figure 1.** Phase portraits from left to right correspond to a stable node, a stable star, and a stable spiral respectively. G = 1, β = 4, and α from left to right is less than $\frac{(\beta-1)^2}{4\beta} + 1$, equal to $\frac{(\beta-1)^2}{4\beta} + 1$, and greater than $\frac{(\beta-1)^2}{4\beta} + 1$.

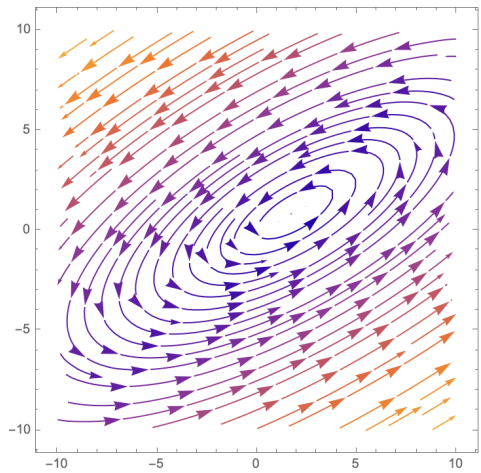

**Figure 2.** Phase portrait displaying a center as the fixed point with G =1, β=1.

## 2.2. Linear Relation between G and I

In the second model, the government spending increases linearly with the national income: $G = G_0 + k * I$, where $k > 0$. Now equation (2) becomes:

$$\dot{C} = \beta \cdot (I - C - G_0 - k * I) \tag{6}$$

Setting and $\dot{C}$ to zero and rearranging terms, I obtain the following equilibrium:

$$(I, C) = \left(\frac{\alpha \cdot G_0}{\alpha(1-k)-1}, \frac{G_0}{\alpha(1-k)-1}\right) \tag{7}$$

and there exists a critical value $k_c = 1 - \frac{1}{\alpha}$. To find an economically sensible equilibrium in the first quadrant, $k$ must be less than $k_c$. The Jacobian matrix and eigenvalues are then computed to determine the stability of the equilibrium:

$$\mathbf{J} = \begin{bmatrix} \frac{\partial \dot{I}}{\partial I} & \frac{\partial \dot{I}}{\partial C} \\ \frac{\partial \dot{C}}{\partial I} & \frac{\partial \dot{C}}{\partial C} \end{bmatrix} \tag{8}$$

$$\lambda_{1,2} = \frac{(1-\beta) \pm \sqrt{(\beta-1)^2 - 4(\alpha(1-k)-1)\beta}}{2} \tag{9}$$

Using the same three alpha values in the first model, Figure 3 displays the phase portraits for $k < k_c$:

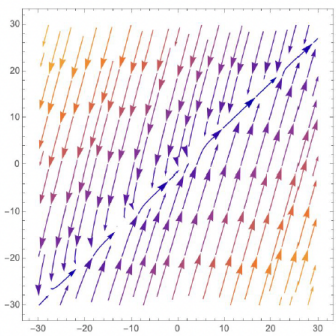
α =1.1

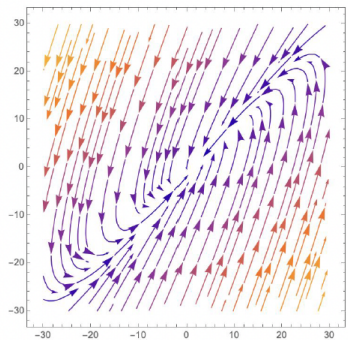
α = 1.5625

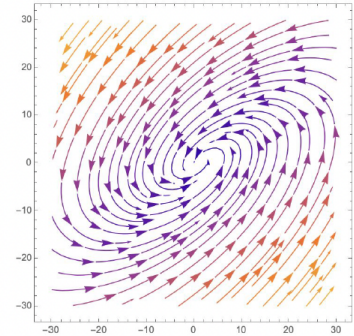
α =5

**Figure 3.** Phase portraits from left to right correspond to a stable node, a stable star, and a stable spiral respectively. $k < k_c$, G = 1, β = 4, and α from left to right is 1.1, 1.5625, and 5.

In the case when $k > k_c$ and β > 0, two eigenvalues will have opposite signs. Accordingly, there exists a saddle node and its stability can be determined by analyzing the correspondent eigenvalues:

$$v_1 = \begin{pmatrix} \frac{\beta+\lambda_1}{\beta(1-k)} \\ 1 \end{pmatrix} \quad v_2 = \begin{pmatrix} \frac{\beta+\lambda_2}{\beta(1-k)} \\ 1 \end{pmatrix} \tag{10}$$

where $v_1$ is the attracting eigendirection and corresponds to the stable manifold; $v_2$ is the repelling eigendirection and corresponds to the unstable manifold. A visualization of this case can be found in Figure 4. The main diagonal from the top left corner to the bottom right corner is the unstable manifold; The antidiagonal from the top right corner to the bottom left corner is the stable manifold.

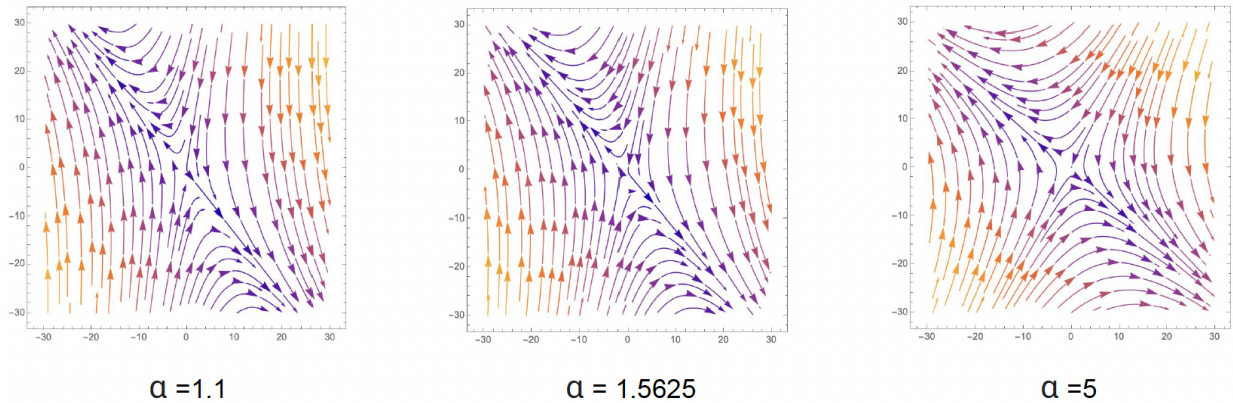

α =1.1     α = 1.5625     α =5

**Figure 4.** Phase portraits showing unstable saddle-shaped trajectories. $k > k_c$, G = 1, β = 4, and α from left to right is 1.1, 1.5625, and 5.

## 2.3. Nonlinear Quadratic Relation between G and I

In the third model, the government spending increases quadratically with the national income: $G = G_0 + k * I^2$, where $k > 0$. Now equation (2) becomes:

$$\dot{C} = \beta \cdot (I - C - G_0 - k * I^2) \qquad (11)$$

Setting and $\dot{C}$ to zero and rearranging terms, I obtain the following equilibrium:

$$(I, C) = (\frac{(1-\alpha) \pm \sqrt{(\alpha-1)^2 - 4k\alpha^2 G_0}}{-2\alpha k}, \frac{(1-\alpha) \pm \sqrt{(\alpha-1)^2 - 4k\alpha^2 G_0}}{-2\alpha^2 k}) \qquad (12)$$

The above fixed point coordinate shows that there exist two fixed points if

$\Delta = \sqrt{(\alpha - 1)^2 - 4k\alpha^2 G_0} > 0$, one fixed point if $\Delta = \sqrt{(\alpha - 1)^2 - 4k\alpha^2 G_0} = 0$,

and no fixed point if $\Delta = \sqrt{(\alpha - 1)^2 - 4k\alpha^2 G_0} < 0$. In other words, there are two fixed points if $G_0 < \frac{(a-1)^2}{4\alpha^2 k}$, one fixed point if $G_0 = \frac{(a-1)^2}{4\alpha^2 k}$, and no fixed point if $G_0 > \frac{(a-1)^2}{4\alpha^2 k}$.

Using the same procedure to compute the Jacobian matrix and eigenvalues as in the first and the second model, I computed the stabilities of fixed points. Figure 5 visualizes fixed points and their trajectories for different $G_0$ values.

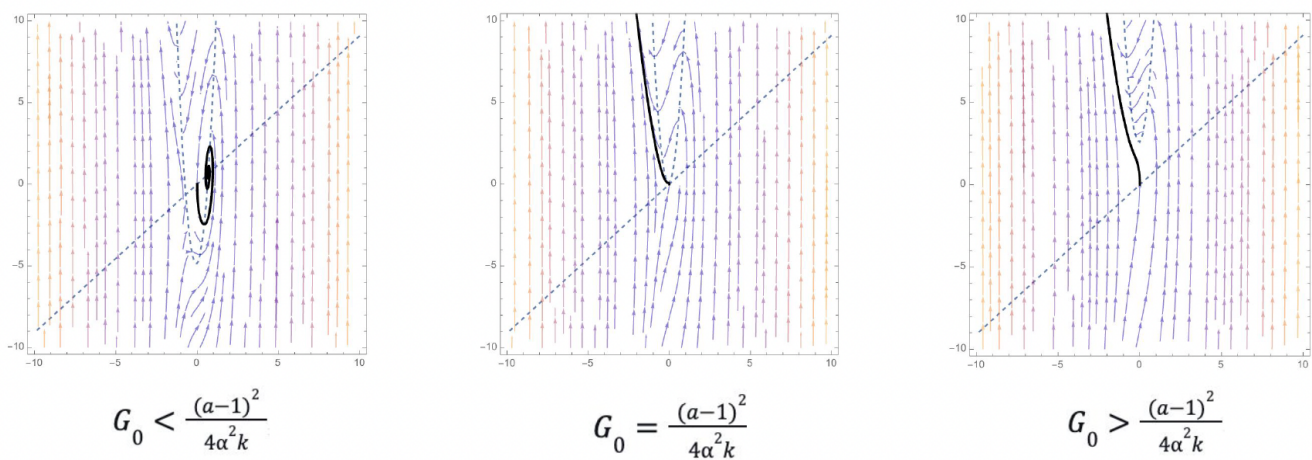

**Figure 5.** Phase portraits displaying fixed points in three different cases. Two dashed lines represent two differential equations of national income and consumer spending and their intersections are the fixed points.

## 3. Results

In the first model, where government spending is a constant, the economic equilibrium is stable as long as $\beta > 1$ because the real parts of the eigenvalues are always negative. Hence, there exists a stable equilibrium state for the economy, which can be represented by the coordinate $(\frac{\alpha \cdot G}{\alpha-1}, \frac{G}{\alpha-1})$. In the limiting case where $\beta = 1$, the eigenvalues are purely imaginary and the solutions have periodic forms. Accordingly, the trajectories are closed curves and the economy is predicted to oscillate.

In the second model, where the government spending increases linearly with the national income, the economic equilibrium is related with a critical value $k_c = 1 - \frac{1}{\alpha}$. If

$k < k_c$, there exists an economically sensible equilibrium in the first quadrant, whereas if $k > k_c$, the equilibrium becomes a saddle node with the main diagonal representing the unstable manifold and the antidiagonal representing the stable manifold. The economy will follow either branch of the unstable manifold depending on the initial condition.

In the third model, where the government spending increases quadratically with the national income, the system can have two, one, or no economic equilibrium depending on how big the initial government spending is. If $G_0$ is less than $\frac{(a-1)^2}{4\alpha^2 k}$, there are two equilibriums in the first quadrant and one of them is a stable spiral. Hence, with the right initial conditions, the economy will approach that equilibrium. If $G_0$ is equal to $\frac{(a-1)^2}{4\alpha^2 k}$, there is a semi-stable equilibrium. If $G_0$ is greater than $\frac{(a-1)^2}{4\alpha^2 k}$, there is no equilibrium and the national income will grow continuously.

## 4. Conclusion

In this paper, I studied the Keynesian cross model of a national economy with a focus on models that describe dynamic relationships between government spending and consumer spending. Using bifurcation diagrams, I analyzed the economically sensible equilibrium states between national income and consumer spending. I discovered that in an economy with constant government spending, its equilibrium is stable if $\beta > 1$ and has a periodic form if $\beta = 1$; In an economy with linearly increasing government spending, its equilibrium is stable if $k > k_c$. Otherwise, the equilibrium becomes a saddle node and the economy will follow either branch of the unstable manifold depending on the initial condition; In an economy with quadratically increasing government spending, there are two equilibriums in the first quadrant and one of them is a stable spiral if the initial government spending $G_0$ is less than $\frac{(a-1)^2}{4\alpha^2 k}$. In the case that $G_0$ is equal to $\frac{(a-1)^2}{4\alpha^2 k}$, there exists a semi-stable equilibrium. Interestingly, when $G_0$ is greater than $\frac{(a-1)^2}{4\alpha^2 k}$, the national income is predicted to grow.

Future researchers can investigate other nonlinear relationships between government spending and national income such as the case when government spending increases cubically with the national income.